\documentclass[twocolumn]{aastex631}
\usepackage{color}

\usepackage{enumitem}
\usepackage{amsmath}
\usepackage{hyperref}
\usepackage{gensymb}
\usepackage{comment}
\usepackage{float}

\usepackage{tabularx} 
\newcommand{\be}{\begin{equation}}
\newcommand{\ee}{\end{equation}}

\newcommand{\lcdm}{\ensuremath{\Lambda\mathrm{CDM}}}

\providecommand{\sorthelp}[1]{}

\begin{document}
\title{Nearly Full-Sky Low-Multipole Cosmic Microwave Background Temperature Anisotropy: \\ 
I. Foreground Cleaned Maps}

\correspondingauthor{Hayley C. Nofi}
\email{hnofi1@jh.edu}

\author[0000-0001-9694-1718]{Hayley~C.~Nofi}
\affiliation{The William H. Miller III Department of Physics and Astronomy,\\ Johns Hopkins University, 3400 N. Charles Street, Baltimore, MD 21218 USA}

\author[0000-0002-2147-2248]{Graeme E. Addison}
\affiliation{The William H. Miller III Department of Physics and Astronomy,\\ Johns Hopkins University, 3400 N. Charles Street, Baltimore, MD 21218 USA}

\author[0000-0001-8839-7206]{Charles L. Bennett}
\affiliation{The William H. Miller III Department of Physics and Astronomy,\\ Johns Hopkins University, 3400 N. Charles Street, Baltimore, MD 21218 USA}

\author[0000-0001-9054-1414]{Laura Herold}
\affiliation{The William H. Miller III Department of Physics and Astronomy,\\ Johns Hopkins University, 3400 N. Charles Street, Baltimore, MD 21218 USA}

\author[0000-0003-3017-3474]{J. L. Weiland}   
\affiliation{The William H. Miller III Department of Physics and Astronomy,\\ Johns Hopkins University, 3400 N. Charles Street, Baltimore, MD 21218 USA}

\shorttitle{}
\shortauthors{Nofi et al} 

\begin{abstract}
Studies of
cosmic microwave background (CMB) are often limited by foreground contamination. Foreground cleaning is performed either in harmonic or pixel space
after data cuts have excluded sky areas of strong contamination. We present a nearly full-sky CMB temperature map with only 1\% of pixels masked. To derive this map, we make use of six full-sky template maps at foreground-dominated frequencies from different experiments smoothed to $1\degree$ and rely on the combination of these weighted maps to trace the morphology of foreground contamination.  We do not impose any spectral index constraints, but
only fit for template amplitudes at each target frequency. We clean WMAP and Planck maps at a set of target frequencies and conduct quality tests at the level of the maps, pixel histograms and power spectra to select four CMB maps that are cleaned with negligible foreground contamination and only 1\% masked pixels
and no inpainting.
We recommend use of these cleaned CMB maps for low multipole ($\ell < 30$) studies.

\end{abstract}

\keywords{\href{http://astrothesaurus.org/uat/322}{Cosmic microwave background radiation (322)}}

\section{Introduction}
\label{sec:intro}

The purpose of this paper is to accurately clean foreground emission from as much of the full sky as possible at the large angular scales $2\le\ell < 30$, including the Galactic plane.  Cosmic microwave background (CMB) observations have been limited by the presence of Galactic foreground emission. Efforts to separate foreground emission from the CMB are based on their differing spectra and/or morphologies. In regions of only moderate contamination corrections are made, but in heavily contaminated regions, especially along the Galactic plane, the data are generally masked (assigned zero weight).

A map of the sky is the most fundamental representation of CMB data. Often a power spectrum is computed from a map with a loss of information. There are two adverse effects of a Galactic plane cut. First, there is the loss of statistical sensitivity due to using only a fraction of the sky. Second, map data are often compressed into spherical harmonics, but these are only orthogonal over the full sky so analysis with a mask leads to mode mixing. 

The Differential Microwave Radiometers (DMR) instrument on the Cosmic Background Explorer (COBE) space mission observed in 3 frequency bands (31.4, 53, and 90 GHz). This enabled the COBE team to discover CMB anisotropy \citep{smoot/etal:1992, bennett/etal:1992b, wright/etal:1992, kogut/etal:1992}, but limited their cosmological analysis to Galactic latitudes $\vert b\vert > 20^\circ$. 

The Wilkinson Microwave Anisotropy Probe (WMAP) space mission team independently confirmed the results of the COBE maps and provided the first precise measurement of all 6 cosmological parameters \citep{bennett/etal:2013}. WMAP observed in 5 frequency bands while still restricting cosmological analysis to Galactic latitudes $\vert b\vert > 20^\circ$ in their initial discovery results. The WMAP team later introduced more tailored sky cuts to replace the simple Galactic latitude cut; however, a substantial sky area was still masked. The WMAP team also introduced an Internal Linear Combination (ILC) technique that provided a full sky map that appeared to be largely cleaned of foregrounds, but the WMAP team warned that there was still significant contamination present \citep{bennett/etal:2003f,bennett/etal:2003e}. 

The Planck mission expanded the frequency sampling to 9 frequency bands (from 30 to 857 GHz) and further improved the precision of the cosmological parameter determination. Most analyses continued to use a large and conservative confidence mask for multipole modes $\ell <30$, while other attempts were made (Commander, NILC, SEVEM, and SMICA) to explore more of the sky~\citep{planckVII/etal:2018}. For example, the Planck common mask excludes 22\% of the sky at HEALPix\footnote{\url{http://healpix.sf.net}} N$_{\rm{side}}=2048$.

The trend towards a growing number of frequency bands over these years highlights both that the foregrounds are complex and difficult to model, and that as instrumental sensitivity to the CMB increased the demands on foreground removal also increased.

Barriers to the removal of foreground emission are due to both its spectral complexity and the use of imperfect 
or insufficient span of emission templates. The CMB spectrum is precisely characterized as a $2.7260\pm 0.0013$ K blackbody based on a combined analysis of WMAP and COBE data \citep{fixsen:2009}. The foreground emission sources  include synchrotron radiation, free-free (thermal bremsstrahlung) emission, thermal dust emission, and Anomalous Microwave Emission (AME). 

The synchrotron radiation spectrum is defined by the relativistic electron energy spectrum at the location of emission. Since the electron spectrum varies by line-of-sight the synchrotron spectral index also varies. Synchrotron emission typically has an antenna temperature spectral index of $\beta_s \sim-3$, but variations are clearly observed across the sky. For example, 
the WMAP haze \citep{planck/intermediate/09:2013}
and Galactic plane supernovae \citep{green:1988, meerkat_snr:2024} have a flatter spectral index of $\beta_s \sim-2.6$. There have been recent detections of variations in the synchrotron spectral index in intensity and polarization in other spiral galaxies \citep{2024AJ....168..138I}. Since other spiral galaxies have a spectral index variation across their plane, we should expect a similar variation across the plane of the Milky Way.

Free-free emission is only weakly dependent on the local electron temperature and frequency. It is expected to have an antenna temperature of spectral index of $\beta_{\rm{ff}} =-2.13$ in the microwave bands \citep{bennett/etal:2003f}. 

Thermal dust emission is dependent on the local radiation field and dust grain size, shape, and composition, with all of these variations integrated along the line-of-sight. This is usually represented as a modified blackbody function with an antenna temperature spectral index  $\beta_d \sim 1.55$ \citep{planck/04:2018}. 

AME is a component of the interstellar medium that appears as excess emission, especially in the frequency range $10-60$ GHz. Although AME has been observed in a variety of studies since its initial detection in 1996 \citep{1996ApJ...464L...5K}, its exact origin remains unknown. It is thought to be emitted by small interstellar dust grains that collide with high-velocity gas, which excites the grains to states of rapid rotation at radio frequencies \citep{1957ApJ...126..480E,silsbee/etal:2011,hensley/draine:2023}. However, no accurate tracer has been discovered, making the removal of AME from CMB maps difficult \citep{dickinson/etal:2018}. Spinning dust emission is an example of AME, but plausible AME theories have a lot of knobs to turn so they can predict a wide range of potential emission behavior. Observations have been used to constrain theories rather than the theories being sufficiently predictive. Observations of individual source spectra indicate a spectral shape similar to a log-normal distribution that typically peaks near $\sim 30$ GHz in flux density units \citep{dickinson/etal:2018}.

In the approximate $20 - 50$ GHz microwave range synchrotron emission, free-free emission, and AME are all described by roughly similar spectral index values making their physical separation challenging. Of these, only free-free emission has a well-specified spectral index. 

To produce maps with a maximal usable sky area, we use template subtraction. Predefined templates are fit independently to clean foregrounds with independent template amplitudes at each target frequency. Instead of foreground-modeled maps, our templates are archival diffuse foreground maps. With this method, we 
do not rely on spectral index assumptions for the individual foregrounds. 

In this paper, we rely on the fact that the CMB and Galactic foreground morphologies are substantially different on large angular scales. We derive a nearly full-sky CMB map with negligible foreground contamination. We focus on the largest scales ($1^\circ$) both because the templates are available at these scales and because it is especially desirable to maximize the usable sky area at these scales. Our aim is to minimize the masking because the larger the excluded region of the sky, the greater the uncertainty in the recovery of low multipole modes. We note that both Planck and WMAP have excellent sensitivity at $1^\circ$, so we clean the maps of both experiments. 

In Section \ref{sec:data} we describe the overall approach and introduce the data criteria. In Section \ref{sec:dataanalysis} we describe our foreground reduction process, discuss masking, and analyze final maps. We summarize and offer brief comments in Section \ref{sec:conclusions}. The focus of this paper is the process used to derive 1\% masked and cleaned CMB maps, the verification of these CMB maps, and ensuring the availability of these maps for the community. In a companion paper \citep{Nofi:2025b}, we use these cleaned CMB maps to determine an accurate low-$l$ power spectrum. In another companion paper \citep{herold2025}, we make use of the nearly full-sky CMB maps to assess large-scale CMB anomalies, removing uncertainties about the effects of masking large portions of the sky.

\section{Data}
\label{sec:data}
\subsection{Concept}

Key aspects of our cleaning include that: (1) we focus only on large angular scales, not the full resolution of the WMAP and Planck data, (2) we do not restrict ourselves to the WMAP and Planck maps. We use external templates (with some requirements listed below), and (3) we clean CMB frequency bands separately, enabling empirical cross-checks of the cleaning between bands.

Our goal is to clean foregrounds from a set of target frequency maps to obtain CMB maps. We judge the success of our cleaning by the similarity of these CMB maps at independent target frequencies. Since earlier template cleaning attempts have had only limited success, it is clear that the cleaning has been over-constrained and more flexibility is needed. We provide this flexibility by not modeling specific emission mechanisms or properties such as spectral indices. Instead, we use foreground templates only for their patterns, not their amplitudes or spectral index values, and then allow the fitting process to use this extra freedom to combine the patterns in whatever proportions best clean the maps.  
Thus, a fundamental concept of our cleaning process is to simultaneously fit numerous foreground templates to the target frequency map to remove large-scale Galactic foregrounds independent of their emission mechanisms.

Although we attempt to represent the morphologies of the main physical components in the selection of templates, {\it we are not concerned with differentiating between the various foreground components or modeling the emission parameters of the foreground component}. We rely instead on the morphologies of multiple template maps to combine in whatever way minimizes the overall foreground contamination at each target CMB frequency.  

\subsection{Data Template Criteria}
\label{datatempcriteria}

\begin{figure*}[ht]
    \includegraphics[width=7in]{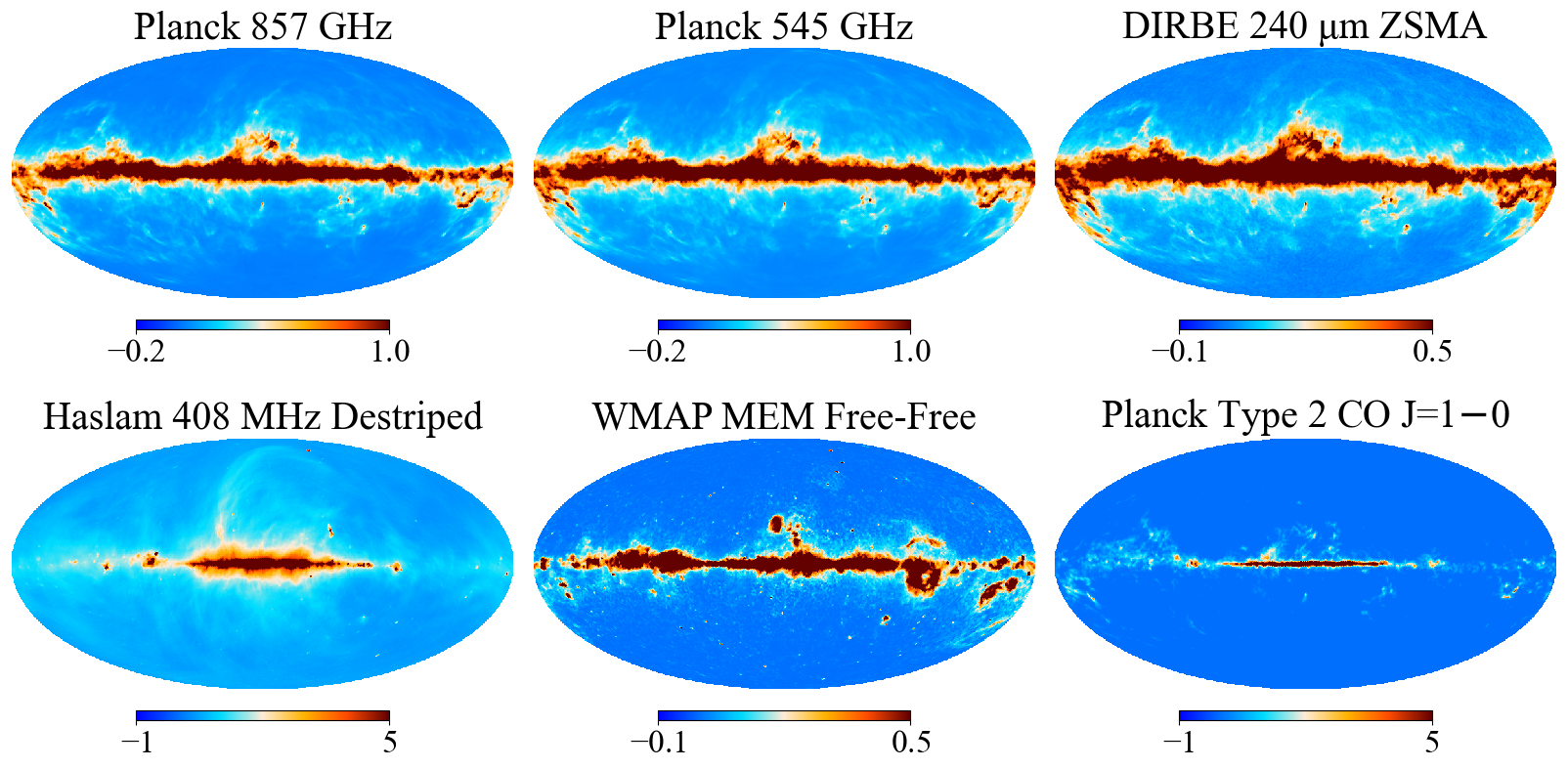}
    \caption{Six template maps used to fit and remove Galactic emission. The maps have been renormalized from their native units to their rms over the sky with the values provided in Table  \ref{tab:coeff_flipped}. 
    Despite the clear similarity of the morphology of Planck 857 and 545 GHz maps there is evidence that using both templates helps cleaning. This may relate to the dust temperature versus spectral index, but we do not address that in this paper.}
    \label{fig:finaltemplates}
\end{figure*}

To determine the data templates used in the cleaning process, we define the following criteria: (1) to create a full-sky CMB map, we require template maps with complete sky coverage, (2) a template map must have a minimum of 1$\degree$ resolution, and (3) a template map must be strongly foreground dominated so that CMB signal removal is negligible during the cleaning process. Based on these criteria, the template candidates are listed in Table \ref{tab:templates} and the final templates used are seen in Figure \ref{fig:finaltemplates}. 

Candidates were rejected as effective templates if they evidenced any of the following characteristics:
(1) their use resulted in notably larger map residuals, 
(2) they did not contribute significantly to foreground removal when used in conjunction with a core set of templates,
(3) they possessed lower signal-to-noise at high Galactic latitudes than equivalent counterparts, or 
(4) a template had missing coverage or identifiable artifacts than could interfere with CMB recovery.
We discuss individual template choices in greater detail in Section~\ref{sub:finaltemp}.

\begin{table*}[t]
    \centering
    \begin{tabular}{lccl}
        \hline
        Map Title & freq or wavelength & Resolution & References and Data File Links  \\\hline

        Haslam 408 MHz destriped & 408 MHz & 56$\arcmin$ &    
        \citealt{remazeilles/etal:2015} \\ 
        & & & 
        haslam408\_ds\_Remazeilles2014.fits\footnote{\url{https://lambda.gsfc.nasa.gov/}}\\
        WMAP MEM free-free & 23 GHz & 1$\degree$ & \citealt{bennett/etal:2013} \\  
        & & &
        wmap\_K\_mem\_freefree\_9yr\_v5.fits$^\mathrm{a}$ \\
        Planck Type 2 CO J=1--0 & 115 GHz & 15$\arcmin$ & \citealt{planck_co:2014} \\  
        & & &
        HFI\_CompMap\_CO-Type2\_2048\_R2.00.fits\footnote{\url{https://pla.esac.esa.int/\#maps}}\\
        Planck 545 GHz & 545 GHz & 4.5$\arcmin$ & 2018 Planck PR3 release \\  
        & & &
        HFI\_SkyMap\_857\_2048\_R3.01\_full.fits$^\mathrm{b}$ \\
        Planck 857 GHz & 857 GHz & 4.2$\arcmin$ & 2018 Planck PR3 release \\ 
        & & &
        HFI\_SkyMap\_857\_2048\_R3.01\_full.fits$^\mathrm{b}$ \\
        DIRBE 240 $\micron$ ZSMA & 240 $\micron$ & 1$\degree$ & \citealt{hauser/etal:1998} \\   
        & & &
        DIRBE\_ZSMA\_10\_1\_256.fits\footnote{\url{https://cade.irap.omp.eu/dokuwiki/doku.php}} \\
        \hline
        Stockert+Villa-Elisa & 1.4 GHz & 35.4$\arcmin$ & 
         \citealt{reichandreich:1986, testori/etal:2001} \\
               
        & & &   STOCKERT+VILLA-ELISA\_1420MHz\_1\_256.fits$^\mathrm{c}$ \\
        HI4PI HI Column Density & 1.42 GHz & 16.2$\arcmin$ & \citealt{HI4PI:2016} \\  
        & & &
        NHI\_HPX.fits$^\mathrm{a}$  \\
        Planck PR2 free-free & 30 GHz & 1$\degree$ & \citealt{planck/10:2016} \\  
        & & &
        COM\_CompMap\_freefree-commander\_0256\_R2.00.fits$^\mathrm{b}$\\
        Planck Revisited CO J=1--0 & 115 GHz & 10$\arcmin$ & \citealt{ghosh/etal:2024} \\  
        & & &
        PlanckRevisited\_CO10\_NSIDE1024\_lmax2048.fits$^\mathrm{a}$\\
        Planck Revisited CO J=2--1 & 230 GHz & 10$\arcmin$ & \citealt{ghosh/etal:2024} \\ 
        & & &
        PlanckRevisited\_CO21\_NSIDE1024\_lmax2048.fits$^\mathrm{a}$\\
        Planck Revisited CO J=3--2 & 345 GHz & 10$\arcmin$ & \citealt{ghosh/etal:2024} \\ 
        & & &
        PlanckRevisited\_CO32\_NSIDE1024\_lmax2048.fits$^\mathrm{a}$\\
        AKARI 160 $\micron$ & 160 $\micron$ & 1.5$\arcmin$ & \citealt{akari:2015} \\ 
        & & &
        AKARI\_160\_1deg\_1\_256.fits$^\mathrm{c}$ \\
        AKARI 140 $\micron$ & 140 $\micron$ & 1.5$\arcmin$ & \citealt{akari:2015} \\ 
        & & &
        AKARI\_WideL\_1deg\_1\_256.fits$^\mathrm{c}$ \\
        DIRBE 140 $\micron$ ZSMA & 140 $\micron$ & 1$\degree$ & \citealt{hauser/etal:1998} \\
        & & &
        DIRBE\_ZSMA\_09\_1\_256.fits$^\mathrm{c}$ \\
        IRIS 100 $\micron$ & 100 $\micron$ & 6$\arcmin$ & \citealt{iris:2005} \\ 
        & & &
        IRIS\_nohole\_4\_1024\_v2.fits$^\mathrm{a}$\\
        WISE 12 $\micron$ & 12 $\micron$ & 12$\arcmin$ & \citealt{wise:2014} \\  
        & & &
        wssa\_sample\_1024-bintable.fits$^\mathrm{a}$   \\
        \hline
        \hline
    \end{tabular}
    \normalsize
    \caption{Full sky maps used as initial templates for foreground removal. The 6 templates above the separation line were the final template set chosen for use in a simultaneous fit to clean foregrounds, leaving behind a candidate CMB map at each target frequency.}
    \label{tab:templates}
\end{table*}

\subsection{Target Maps}
The target maps for cleaning are archival WMAP and Planck maps from the 2013 WMAP DR5 (9-year) release\footnote{\url{https://lambda.gsfc.nasa.gov/product/wmap/current/index.html}}
and the 2018 Planck PR3 release\footnote{\url{https://pla.esac.esa.int/\#maps}}. 
For WMAP we tried to clean the 23 GHz K, 33 GHz Ka, 41 GHz Q, 61 GHz V, and 94 GHz W band maps and for Planck we tried to clean the 30, 44, 70, 100, 143, and 217 GHz bands. We found, not surprisingly, that the target frequencies near the foreground-to-CMB signal minimum were most effectively cleaned. Therefore, we limit the main results of this paper to the 4 bands in the 70-143 GHz range of WMAP and Planck.

\section{Data Analysis}
\label{sec:dataanalysis}
\subsection{Overview}
To remove the Galactic foregrounds from the WMAP and Planck target band maps, we implement a modified template subtraction technique. Since we fit over large sky regions with sufficiently high signal-to-noise foreground maps there is only a low probability of chance correlations between the CMB and foreground templates\footnote{We examine the impact of CMB-foreground chance correlation in detail in Section~3.2 of \cite{Nofi:2025b}.}. Therefore, the CMB is not much altered by the template foreground removal fits. We characterize the CMB target map at each frequency, $\nu$, as the observed map minus a superposition of foreground morphology templates:
 
\begin{equation}
   T_{\nu}^{\mathrm{CMB}} = T_{\nu}^{\mathrm{obs}} - (\Sigma a_{i,\nu}T_{i} + C_{\nu}),
\label{equ:temp}
\end{equation}

\noindent where $T_{\nu}^{\mathrm{obs}}$ is the frequency band map, $T_{\nu}^{\mathrm{CMB}}$ is the CMB in that frequency band, and $a_{i,\nu}$ are fit coefficients for each template map $T_{i}$ at frequency $\nu$. A constant factor $C_{\nu}$ is added to absorb monopole normalization differences between the maps. We use the SciPy \texttt{curvefit} non-linear least squares method to fit for the coefficient values \citep{virtanen/etal:2020} by minimizing the differences between the publicly delivered as-observed frequency band map and the combination of foreground templates with flat weights. Based on Planck FFP10 simulations\footnote{\url{wiki.cosmos.esa.int/planck-legacy-archive/index.php?title=Simulation_data}} and WMAP documentation\footnote{\url{lambda.gsfc.nasa.gov/product/wmap/dr5/pub_papers/nineyear/supplement/WMAP_supplement.pdf}}, the mean noise rms values are $\sigma^{70}_{noise} = 2.5$ $\mu$K, $\sigma^{94}_{noise} = 5.0$ $\mu$K, $\sigma^{100}_{noise} = 0.79$ $\mu$K, and $\sigma^{143}_{noise} = 0.65$ $\mu$K. We fit over large sky areas where the noise is much smaller than the signals from the CMB (70~$\mu$K rms in the Planck \lcdm\ model) and the foregrounds. We therefore use uniform weighting in the Equation \ref{equ:temp} fit, since this is a signal-dominated large-scale analysis and the fit is insensitive to detailed noise properties.

In Section \ref{sub:dataprep} we describe the preparation of templates and data maps, including smoothing to a common resolution, handling of the kinematic quadrupole and Zodiacal emission, and masking of a small portion of the sky close to the Galactic center. In Section \ref{sub:initialclean} we describe the results of our initial template-based cleaning, with no additional mask applied. In Section \ref{sub:finaltemp} we review the choice of foreground templates and our decision to use six of the available templates in the final analysis. In Section \ref{sub:masks} we compare the cleaned maps, finding evidence for small foreground residuals close to the Galactic plane, and show that masking 1\% of the sky is sufficient to largely eliminate them. In Section \ref{sub:finalclean} we repeat the template cleaning excluding these 1\% of pixels from the beginning so they do not impact the template coefficients. We present some additional checks and visualization of the final cleaned maps in Section \ref{sub:analysis}.

\subsection{Data Preparation}
\label{sub:dataprep}
We prepare the templates and frequency bands for cleaning by smoothing to 1\degree\  (FWHM) and then downgrading to a common HEALPix resolution of $N_\mathrm{side}= 128$. The templates are normalized to their rms over the full sky for the fitting process. This normalization avoids extreme numerical values but has no impact on results. The $a_{i}$ scaling is not constrained and depends on the selected normalization of the templates. 

The kinematic quadrupole is due to our proper motion with respect to the CMB rest frame. It was removed from the delivered Planck band maps but not from the delivered WMAP band maps \citep{Planck:2015HFI,Planck:2015LFI,hinshaw/etal:2009}. We remove the kinematic quadrupole from the WMAP band maps to consistently focus on cosmological signals.

Interplanetary dust (IPD) in our solar system both scatters solar radiation in the infrared and re-emits thermal energy from the Sun into submillimeter wavelengths. 
This latter signal is referred to as zodiacal emission.
Zodiacal emission was removed by the Planck Collaboration in their PR3 maps, using a parametric model \citep{planck_zodi:2014, dirbe_zodi:1998}. However, our initial cleaning residuals at 100 and 143 GHz showed a faint but distinctive residual 
signature appearing as a pair of bands centered at
ecliptic latitudes $\pm 10\degree$. This is associated with the ``Band 1'' zodiacal model component, usually ascribed
to dust located in specific asteroidal family orbits. The
zodiacal light residuals indicated that the Planck model slightly over-corrected at these frequencies. Once we identified this issue, we sought to correct the 100 and 143 GHz maps for the zodiacal
oversubtraction.   We determine the amplitude of the zodiacal residual in the 100 and 143 GHz foreground cleaned maps by performing a separate fitting step in which we first subtract a zodi-free estimate of the CMB and then fit a zodiacal band template to these two CMB and foreground reduced maps.
We use the ZodiPy package
\citep{zodipy:2022} to generate Planck-specific maps of the $\pm10^\circ$  band pair, evaluated for four individual days equally spaced over
a year. The template itself is the unweighted average of these four maps, normalized using the rms method described previously.  
The zodiacal band template fit amplitudes for the 100 and 143 GHz maps agreed within the uncertainties, so we adopted the same 3.8 $\mu$K amplitude for both.
This is a low-level
correction compared to other signals in the maps.
New foreground cleaned fits using these zodi-corrected 100 and 143 GHz maps as inputs no longer have any zodiacal signatures.
Our zodi correction template will be made public.

The Galactic center Sagittarius A region is unusual in three ways relevant to our cleaning: (1) it is extremely bright, (2) it has an unusual spectrum that peaks near 90 GHz, and (3) it is highly time-variable \citep{Witzel:2021}. 
Before our initial cleaning, we excluded a region within 1$\degree$ of Sagittarius A or about 0.02$\%$ of the pixels in the sky. This was to prevent this especially bright and atypical region from affecting fits. 

\subsection{Initial Cleaning Process}
\label{sub:initialclean}
Cleaning of the brightest foreground emission, near the Galactic plane, is the most demanding part of the model fitting in terms of required accuracy. Other parts of the sky, with much lower foreground levels, need not be as exacting to be effective.  
Therefore, we fit the Galactic foreground emission templates only over the portion of sky where foregrounds are brightest, along the Galactic plane. We chose Galactic latitudes $\pm 10 \degree$. The template coefficients are then applied across the entire sky for subtraction. We are able to apply these coefficients to the full sky because at higher latitudes, the foreground signal is much weaker, and this allows greater tolerance for foreground removal imperfections. 

A residual monopole and dipole due to calibration differences can still exist in the maps and we remove these after Galactic foreground removal so that the Galactic signal does not interfere with the residual fit. We execute the subsequent fit using \texttt{curvefit} applied to the full-sky.

\begin{figure}
    \includegraphics[width=3in]{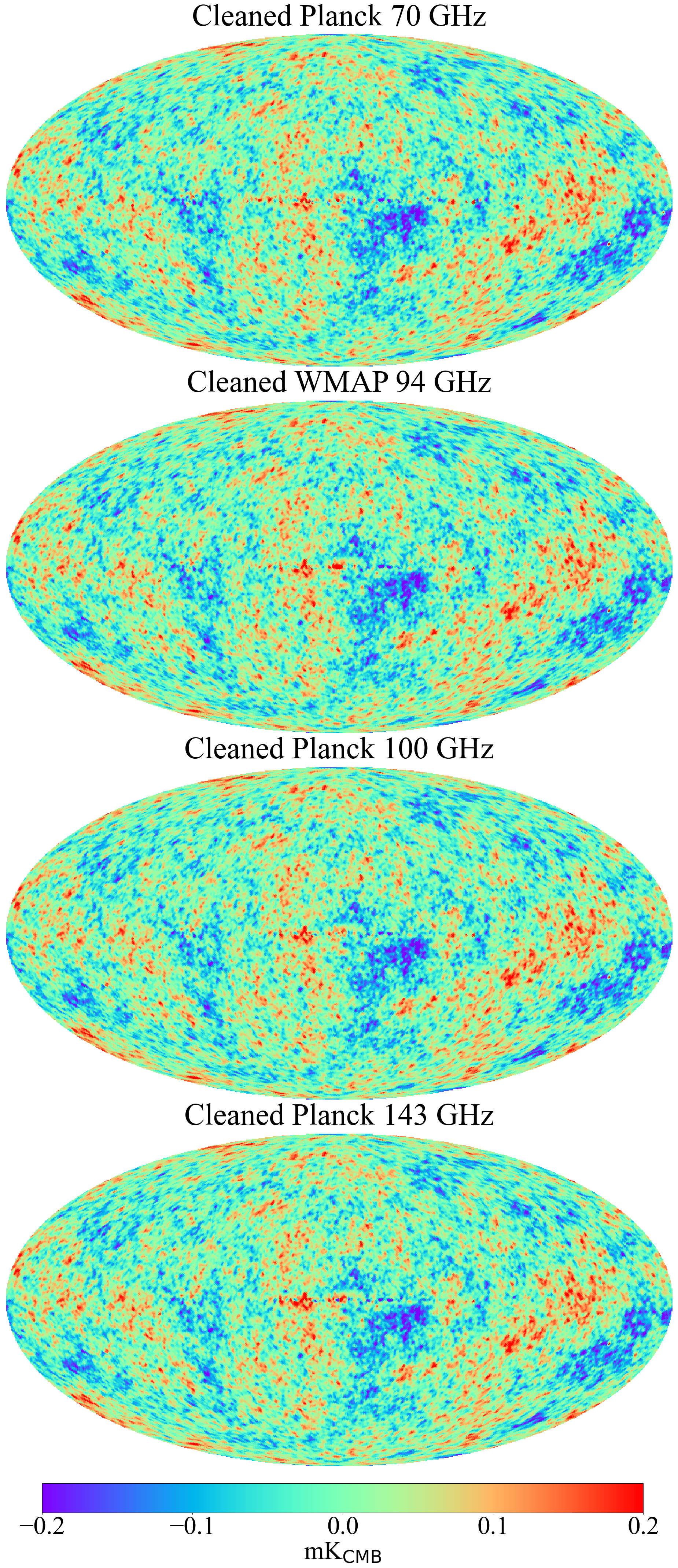}
    \caption{Full sky projections in Galactic coordinates of the best cleaned maps in thermodynamic temperature after the initial cleaning. The maps are well-cleaned over most of the sky but with some foreground remaining along the Galactic plane, especially in the general vicinity of the Galactic center, which is at the center of the maps. The visual similarity of the maps is obvious.}
    \label{fig:fullsky}
\end{figure}

Based on map differences, power spectra differences, histogram comparisons, and other statistical comparisons, we find that the best-cleaned CMB maps are the four at target frequencies of 70, 94, 100, and 143 GHz shown in Figure \ref{fig:fullsky} after the initial cleaning. This is likely because these bands are near the spectral minimum of the Galactic-to-CMB antenna temperature ratio, but it is also probably affected by the degree to which morphological complexities are represented by available templates at each frequency. For example, there are no high-quality templates that trace the AME or allow variation of the synchrotron spectral index.

\subsection{Finalizing Template Choices}
\label{sub:finaltemp}
With the templates established, we sought to determine which linear combination of spatial template patterns best matches the observed Galactic emission for each frequency independently without restrictions on the spectral index. See Section \ref{sub:masks} for examples of the quantitative tests we used to address cleaning quality. During the fitting process, we found that some templates were far more effective than others in removing Galactic signals. Once these most effective templates were applied, the use of additional secondary templates did not significantly reduce residuals. Below, we offer brief commentary on the templates that were and were not preferred by the fits.

\begin{figure*}
    \includegraphics[width=7in]{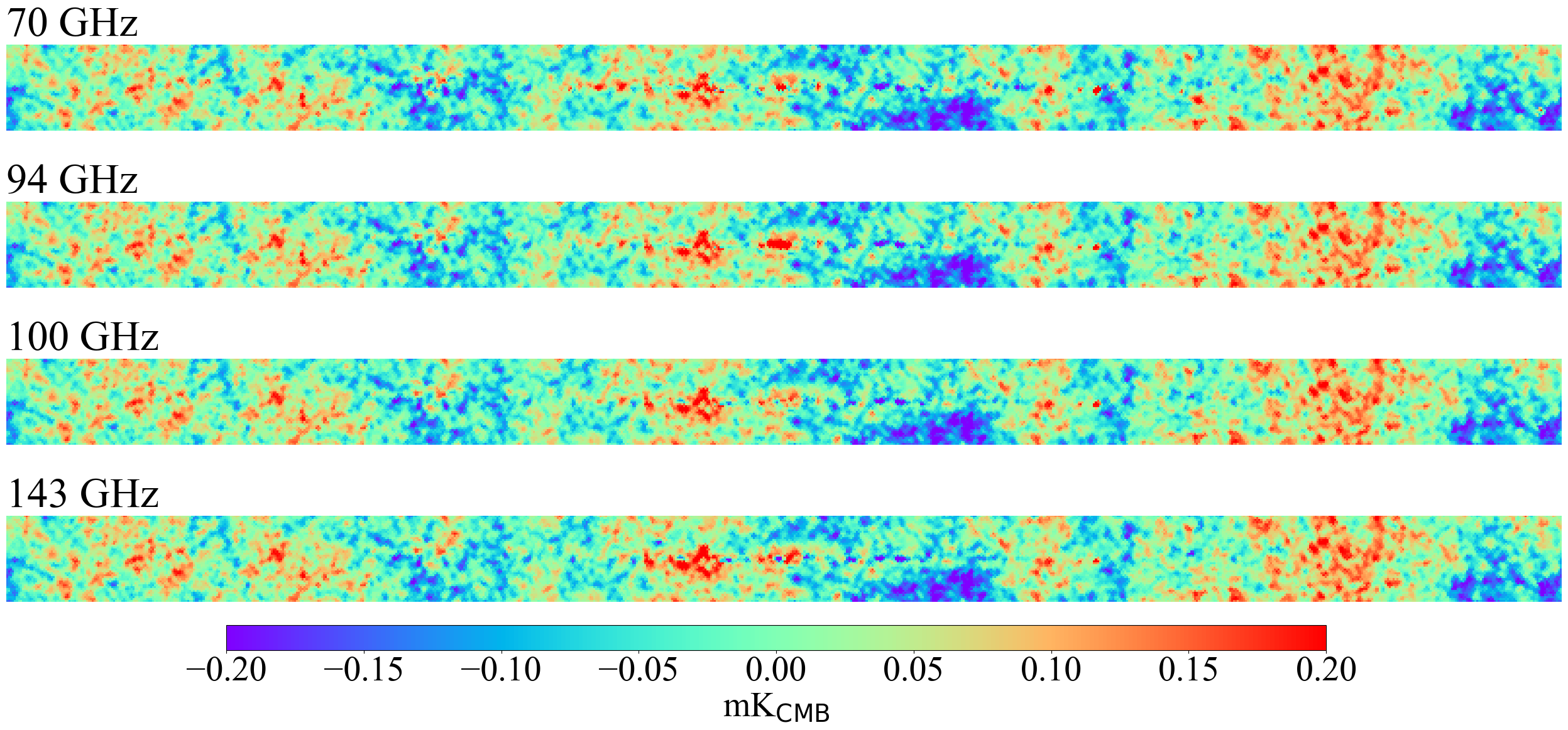}
    \caption{Projection of the best cleaned maps in Galactic coordinates of Galactic latitude $\pm10 \degree$ after the initial cleaning. The top map is the cleaned Planck PR3 70 GHz band, the second map is the cleaned WMAP 94 GHz W band, the third map is the cleaned Planck PR3 100 GHz band, and the bottom map is the cleaned Planck PR3 143 GHz band. Some foreground residuals along the Galactic plane are visible, but greatly reduced compared to previous efforts and the substantial similarity of these strips to one another is apparent.}
    \label{fig:glat}
\end{figure*}

The cleaning process does not appreciably improve with Wide-field Infrared Survey Explorer (WISE) 12 $\micron$, Improved Reprocessing of the Infrared Astronomical Satellite (IRAS) Survey (IRIS) 100 $\micron$,  AKARI 140 $\micron$, AKARI 160 $\micron$, and HI4PI included in the fit with the six templates producing the best cleaned maps. The DIRBE 140 $\micron$ template behaved similarly to the DIRBE 240 $\micron$ template when included in the fit with the six selected templates. Substituting DIRBE 140 $\micron$ for DIRBE 240 $\micron$ in the six-template fit yielded nearly identical results. We opt for DIRBE 240 $\micron$ due to the better signal-to-noise at higher latitudes \citep{hauser/etal:1998}. These discarded templates might have helped trace thermal dust and/or Anomalous Microwave Emission (AME), but did not effectively do so as well as the six selected templates.

Early cleaning tests at low frequencies (23 - 70 GHz) indicated a marked preference for the Haslam 408 MHz template
over that of the 1.4 GHz Stockert + Villa Elisa map.  This, along with the large-scale morphological differences
between these two surveys (seen both visually and in e.g. \citealt{weiland/etal:2022}), and the better  
signal-to-noise at high latitudes in the Haslam map, acted to eliminate the 1.4 GHz survey from our core template set.

\begin{figure*}
    \includegraphics[width=7in]{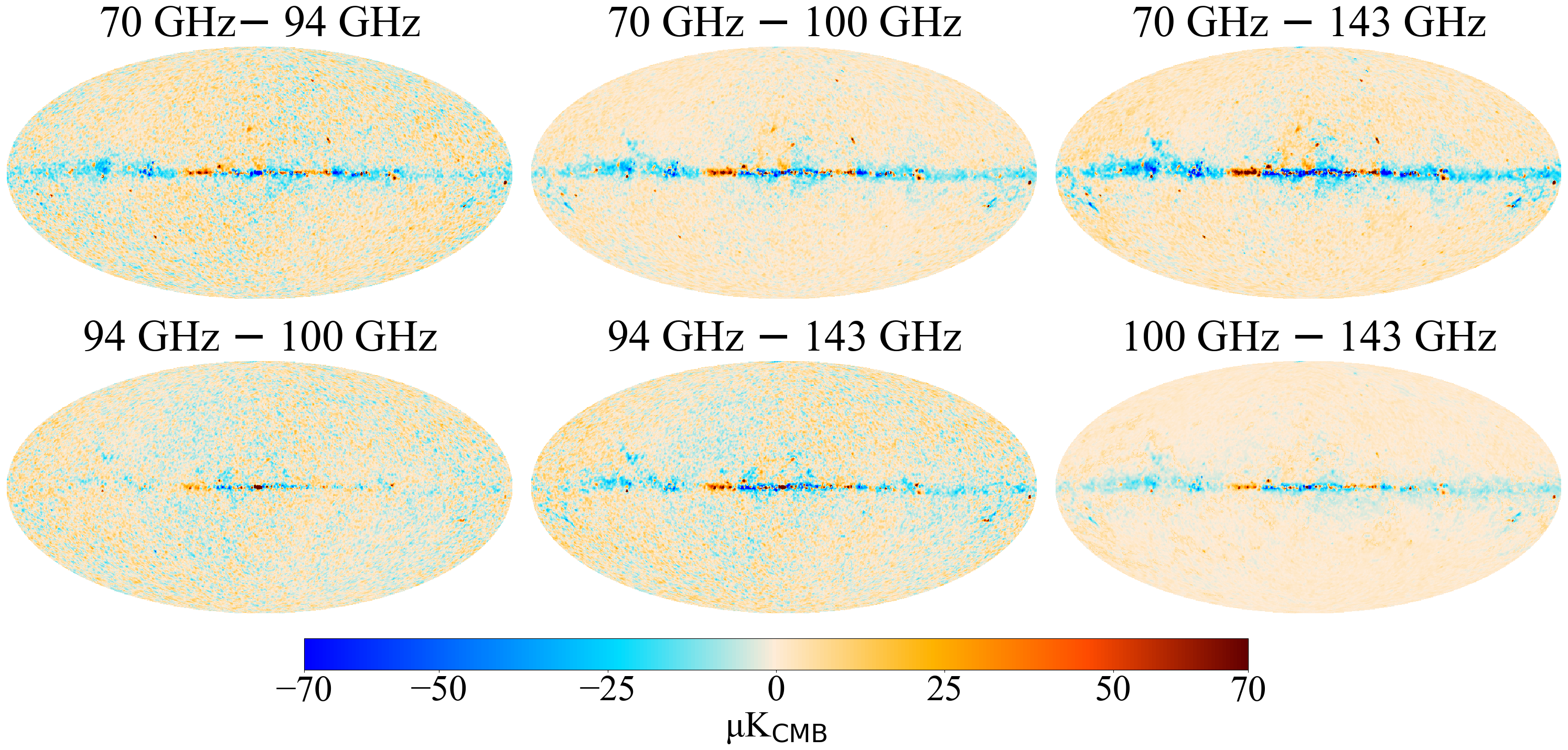}
    \caption{Full-sky differences between the best-cleaned maps for the initial cleaning, in Galactic coordinates. Since the maps are in thermodynamic temperature units the CMB should cancel, aside from gain or other instrumental differences. The full color stretch corresponds to $\pm 1\sigma$ of the cosmic variance at the map resolution. We note that a small difference is not the same as being correct since common-mode flaws will cancel in the differences. In addition to CMB diffuse foreground cleaning errors, time-variable sources and systematic errors may also produce a residual. Our overall conclusion is that these four maps are all good representations of the CMB sky at $1\degree$ resolution with foreground removal errors below the cosmic variance level.}
    \label{fig:differencesr1}
\end{figure*}

Free-free emission has a unique morphology, so a template for it is necessary to clean foregrounds.
However, there is no directly observed map where free-free emission dominates because, although it is present over a wide range of frequencies, there is no frequency range where it greatly outshines all other emission sources. 
Ionized regions that generate free-free emission also generate H$\alpha$ spectral lines. 
However, deriving a free-free template from H$\alpha$ observations \citep{finkbeiner:2003, haffner/etal:2010}  requires corrections for dust extinction and scattering, which are highly uncertain near the plane. Corrected H$\alpha$ maps have been used as free-free templates (see e.g. 
\citealt{harper/etal:2022}) at high Galactic latitudes where the corrections are smaller.
Both free-free templates in Table \ref{tab:templates} (from WMAP and Planck) are derived quantities rather than direct observations. These free-free templates were derived using parametric fits to multiple frequency bands that include a CMB component. 
WMAP removes an estimate of the CMB and Planck fits for it as a separate component, but there may be some low-level CMB remnant in these templates, although greatly subdominant to free-free emission.  
The WMAP Maximum Entropy Method (MEM) free-free sky map \citep{bennett/etal:2013} is based on H$\alpha$ observations as a prior. For high optical depth H$\alpha$ lines of sight (generally at low Galactic latitudes), the WMAP-observed free-free intensity is strong enough so that the MEM results are not strongly informed by the H$\alpha$ prior. 
However, use of the prior limits both the
high-latitude noise and potential for contamination from other signals at higher latitudes where the free-free 
signal is weakest.
The use of the WMAP free-free map should have a negligible effect on the target map CMB signal. The Planck template does not use an H$\alpha$ prior, so it could be slightly more influenced by the CMB at high latitudes. Thus, we chose the WMAP free-free template, although we found no great difference in the results between the two options.

The CO J=1-0 rotational transition at 115 GHz falls in the Planck 100 GHz bandpass. The CO J=2-1 transition at 231 GHz is in the Planck 217 GHz bandpass, and the J=3-2 transition at 346 GHz is in the 353 GHz bandpass. The analog transitions of $^{13}$CO J=1-0 at 110.2 GHz, J = 2-1 at 220.40 GHz and J = 3-2 at 330.6 GHz also lie in these Planck bandpasses.

To allow for a CO emission component, we tested four different Planck CO template maps, the Planck Type 2 CO=1-0 map \citep{planck_co:2014} and the \cite{ghosh/etal:2024} Planck Revisited maps of CO J = 1-0, J = 2-1 and J = 3-2.  Since the $^{13}$CO emission is correlated and fainter than the analog $^{12}$CO maps, we did not include different templates for these transitions. 
When used in conjunction with the other five core templates, fits using the Type 2 CO J=1-0 map produced lower residuals within the plane than the other CO templates.  The Planck Revisited CO maps have less noise
at high latitudes, however.  We compensate for the higher noise in the Type 2 CO map by applying a thresholding cut such
that signals \boldmath$\leq$0.5 K km s$^{-1}$ are set to zero in the template.

\begin{figure*}[ht!]
    \includegraphics[width=7in]{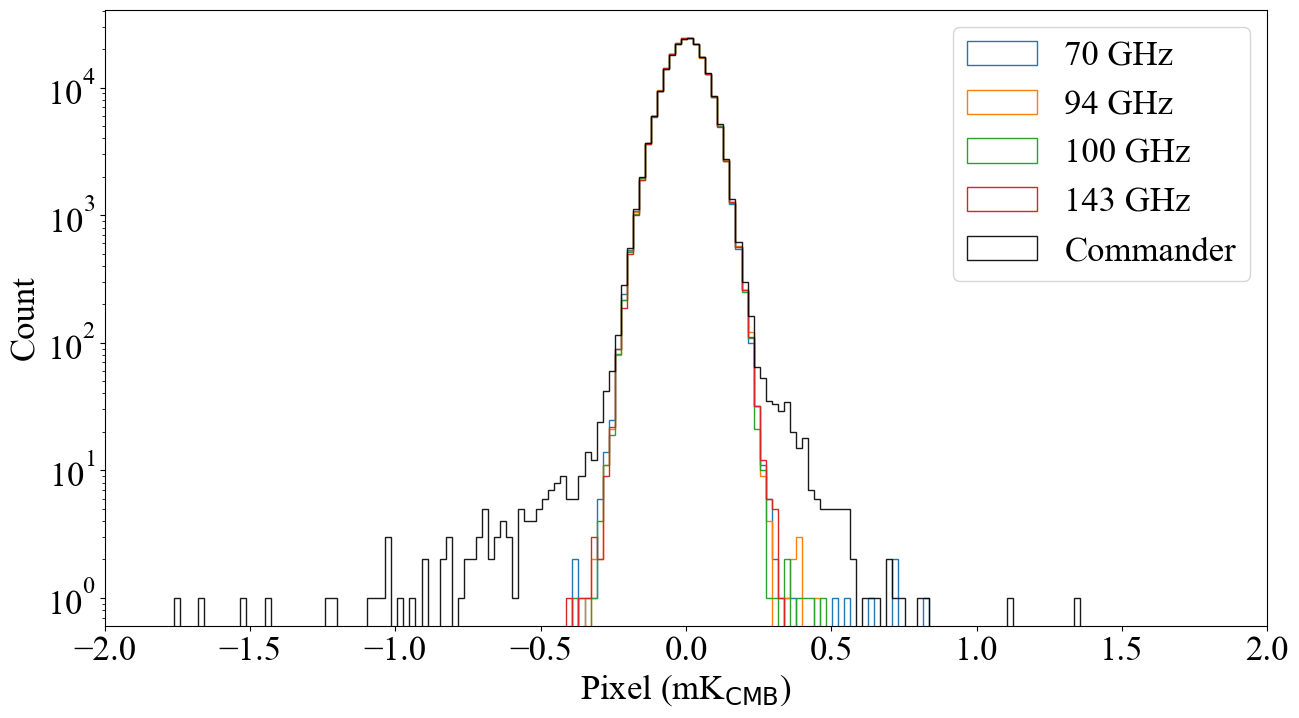}
    \caption{Pixel histograms of the four cleaned maps from the initial cleaning using the Sagittarius A mask. We have included the PR3 Planck Commander CMB map (provided over the full sky with no pixel inpainting) as a comparison. We acknowledge that the Planck Collaboration recommends the use of a confidence mask for any cosmological analysis of this map. The histograms of the four cleaned maps from this work are generally in good agreement, although there are a small number of outlier pixels from the otherwise-symmetrical distributions that are not consistent across frequency, indicating residual foreground contamination and motivating a small amount of additional masking (see text).}
    \label{fig:hist}
\end{figure*}

There are three templates (Planck 857 GHz, Planck 545 GHz, and COBE/DIRBE 240 $\micron$) that all contribute significantly to tracing the thermal dust, but also presumably trace AME to some extent. Haslam 408 MHz traces the synchrotron, and the WMAP MEM free-free map nominally traces ionized thermal emission. We note that there is no direct template capability to adjust for synchrotron spectral index variations. Although spectral behavior is not imposed, expectations can be checked against the derived spectral behavior. 

We decided to limit the number of templates used in the CMB cleaning for several reasons. The main reason is that once the foregrounds are well-removed, there is a danger that additional templates may introduce template systematic errors with no cleaning benefit. Also, a much larger number of templates could start to conspire to cancel the CMB.  The fundamental idea of the fits is that foregrounds have a significantly different morphology than the CMB, but this relies on fitting a limited number of templates over a large sky area.

\subsection{Contamination Analysis and Masks}
\label{sub:masks}

An expanded view of the Galactic plane region of the four best maps after the initial cleaning is compared in Figure \ref{fig:glat}. Although the maps are similar to the eye, we can probe more deeply by taking map differences in thermodynamic temperature units, so the CMB emission cancels as seen in Figure \ref{fig:differencesr1}. The color bar stretch has been set to \unboldmath$\pm 70 \mu{\rm K}$ to correspond to $\pm 1\sigma$ of the CMB anisotropy. Here we see cleaning residuals, however, the great majority of sky pixels have foreground contamination that is much less than the CMB rms.

The cleaning residuals can also be seen in the histograms of the pixel values in the four target CMB maps in comparison with the Planck 2018 PR3 \texttt{Commander} map\footnote{COM\_CMB\_IQU-commander\_2048\_R3.00\_full.fits} \citep{planck/04:2018}, shown in Figure \ref{fig:hist}. It is clear that there are a number of pixels that were not ideally cleaned, as indicated by the frequency differences in Figure \ref{fig:differencesr1}.
In the case of the \texttt{Commander} CMB map, this is not surprising, since this product was not intended to represent the CMB with confidence near the Galactic plane.  In our analysis, however, the few outliers are of concern, since our goal is to maximize the usable sky fraction.

We conclude that we need to mask some pixels in the CMB maps that are outliers in the histogram and that did not clean well enough. We have four candidate CMB maps and each pixel should agree within the statistical errors between the four maps.  We use this as a guide for masking. We construct a standard deviation based mask where we compute a standard deviation for every pixel across the sky in the four cleaned CMB maps after the initial cleaning. We then take the per-pixel standard deviations and include the pixels from the Sagittarius A region in the mask. We mask based on a top percentage level of the ranked standard deviations. We did this for a range of percentages of masked pixels, with the minimum at 0.5\% masked pixels up to 10\% or more of masked pixels. (In the end, all but one pixel from the Sagittarius A region would have been automatically added to the 1\% and higher masks.)

\begin{figure}[hbt!]
    \includegraphics[width=3.2in]{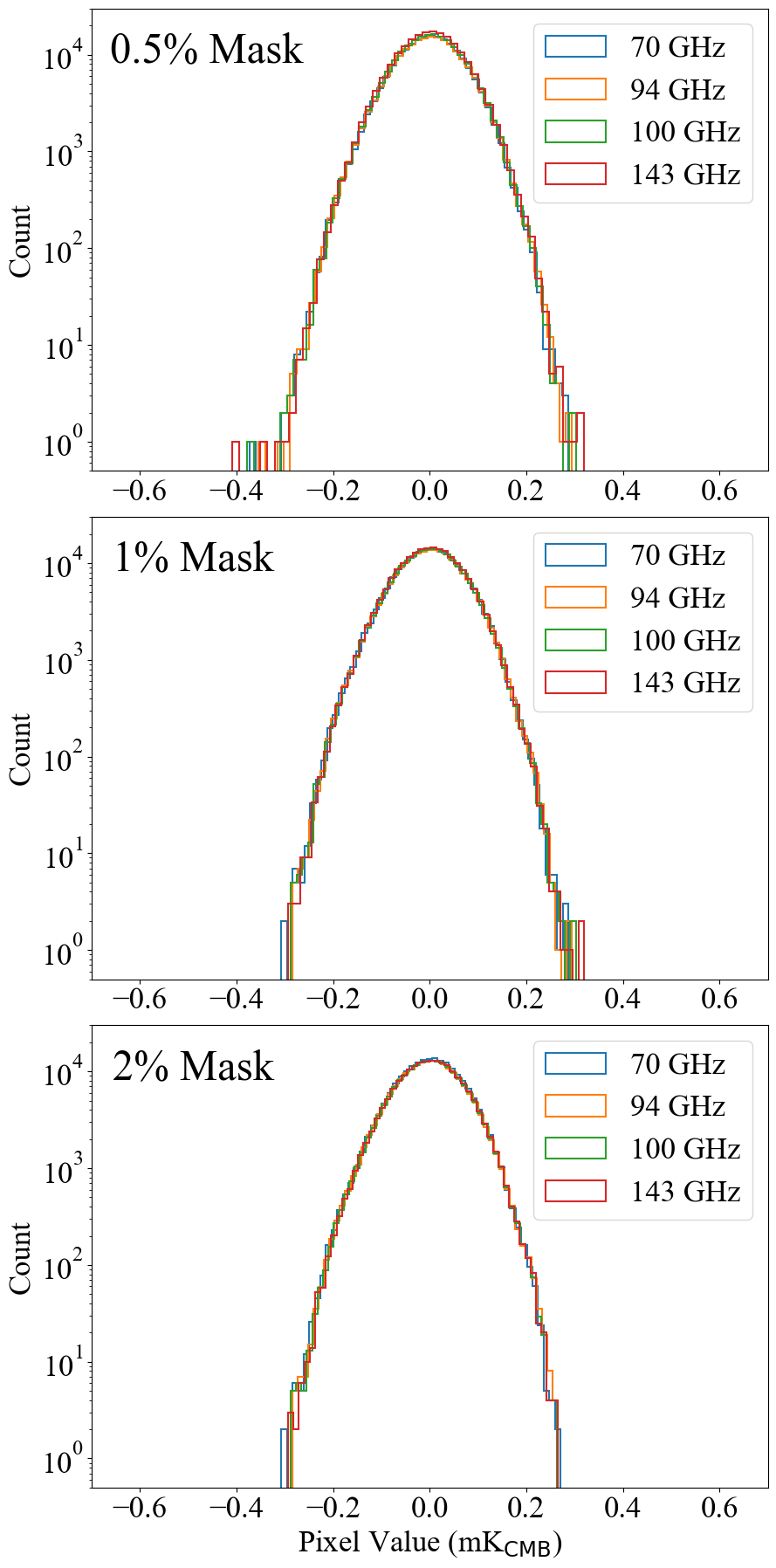}
    \caption{ 
    Pixel histograms masking 0.5\% (top), 1\%  (middle), and 2\%  (bottom) of the map pixels. The masked pixels are those with the highest standard deviation between the 70, 94, 100, and 143 GHz cleaned CMB maps from the initial cleaning. We observe that there is no dramatic change between these histograms. The 1\% mask is likely slightly better than the 0.5\% mask and any differences with the 2\% mask are minimal. 
    The exact masking percentage selection is not critical. We adopt the 1\% mask.}
    \label{fig:stdvhist}
\end{figure}

The three histograms in Figure \ref{fig:stdvhist} show the pixel values for
three different masking levels (0.5$\%$, 1$\%$, and 2$\%$ of pixels masked).
As can be seen, the pixel-value distribution is completely
dominated by CMB fluctuations, and the outlier pixels shown
in Figure \ref{fig:hist} are removed. We recommend the 1\% masked CMB map, but we also conclude that the purity of the CMB map is not sensitive to this specific choice of masking, as seen in the histograms. 
Further quality tests using power spectra of the final cleaned maps, discussed in Section~\ref{sub:analysis}, also support this conclusion.

\begin{figure}[ht]
    \includegraphics[width=3.2in]{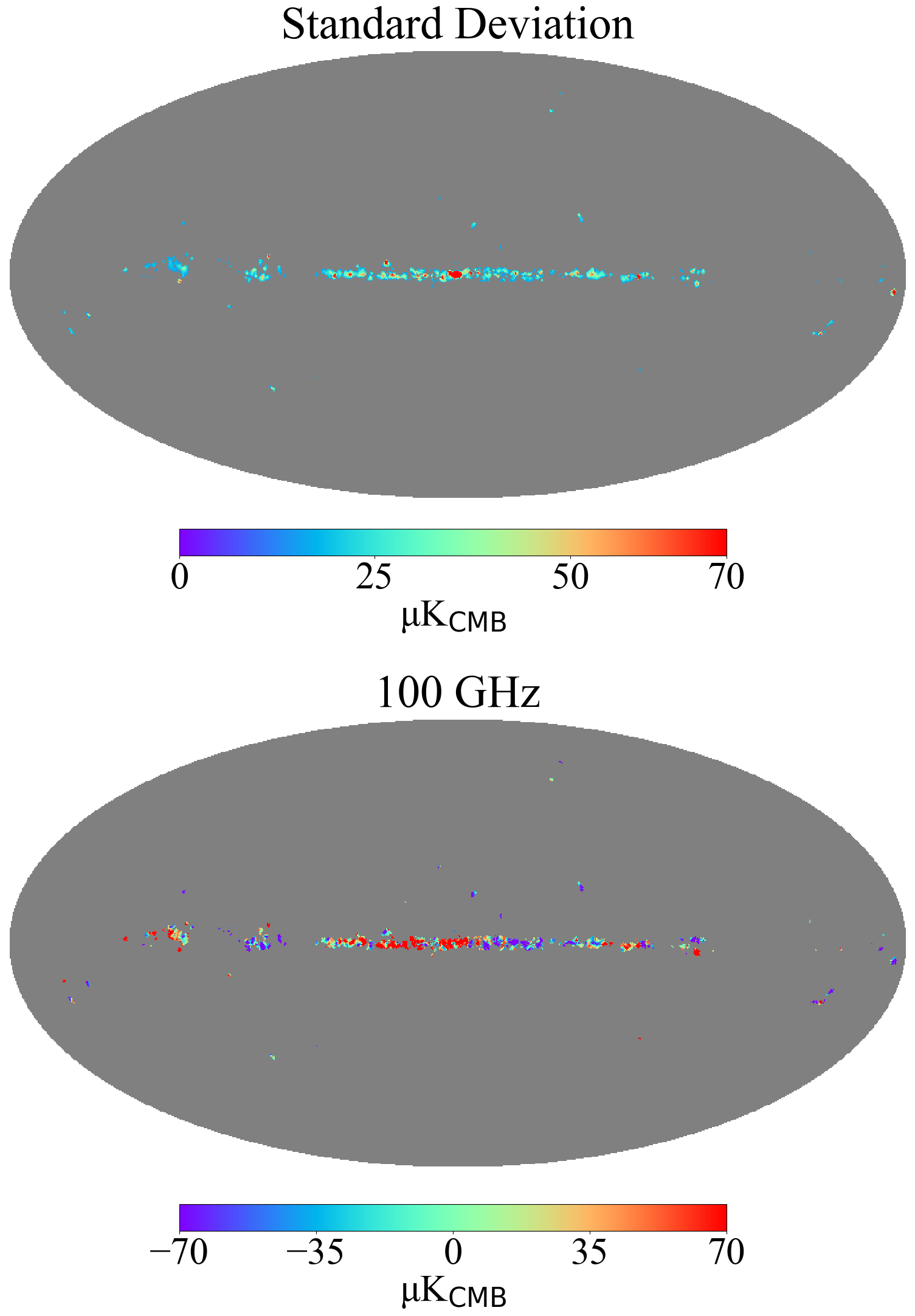}
    \caption{Two views of the 1\% mask region. 
    Top: The absolute value of the standard deviation of the four best cleaned maps from the initial cleaning where the $1\%$ mask applies. Most of the masked pixels are a fraction of the CMB variance in the cut region, while a small number are comparable or greater.  The full color bar stretch is set to the $1\sigma$ CMB rms level at this resolution.   
    Bottom: the temperature of the cleaned 100 GHz map from the initial cleaning where the $1\%$ mask applies. While temperature of the bright pixels in the Galactic plane are substantially reduced in the cleaning process, some pixels should still be masked. 
    } 
    \label{fig:mask_views}
\end{figure}

We illustrate the 1\% mask in Figure \ref{fig:mask_views}. The top map shows the masked pixels colored with the standard deviations of the four CMB maps from the initial cleaning. We have set the color bar stretch to show $1\sigma$ of the CMB fluctuations at the map resolution.  

Most of the pixels masked in the 1\% cut have amplitudes below the CMB rms, but corresponded to a relatively higher standard deviation between the four cleaned maps. While cleaning errors may dominate these pixels, it is also possible that other effects contribute, such as source variability and various systematic errors related to beam uncertainties, ADC nonlinearity, calibration, etc.

\begin{figure}[ht!]
    \includegraphics[width=3in]{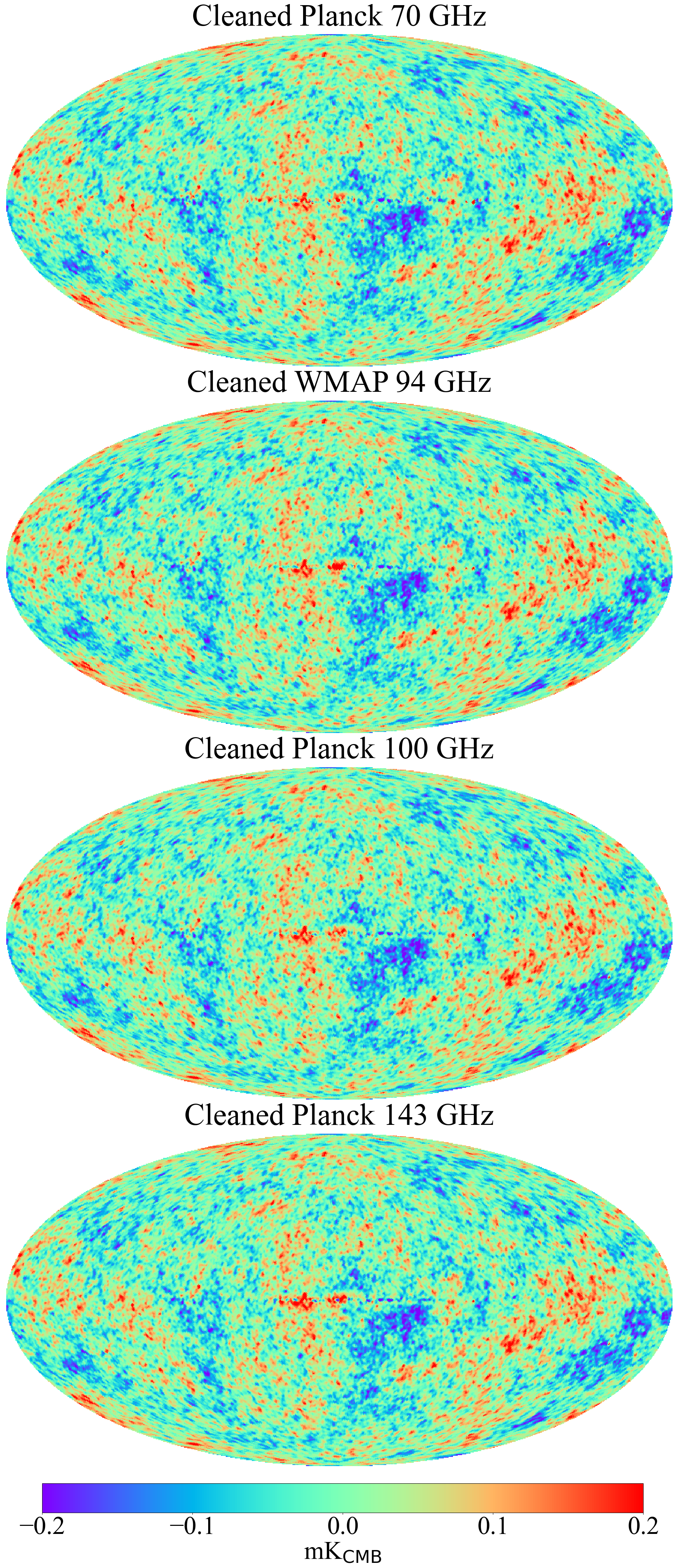}
    \caption{The best-cleaned CMB maps after we have re-fit the template coefficients excluding the 1\% of masked pixels. See the caption of Fig. \ref{fig:fullsky} for details. The results shown here are dominated by CMB fluctuations over 99\% of the sky for all four target bands.}
    \label{fig:fullskyr2}
\end{figure}

\begin{figure}[ht!]
    \includegraphics[width=3.3in]{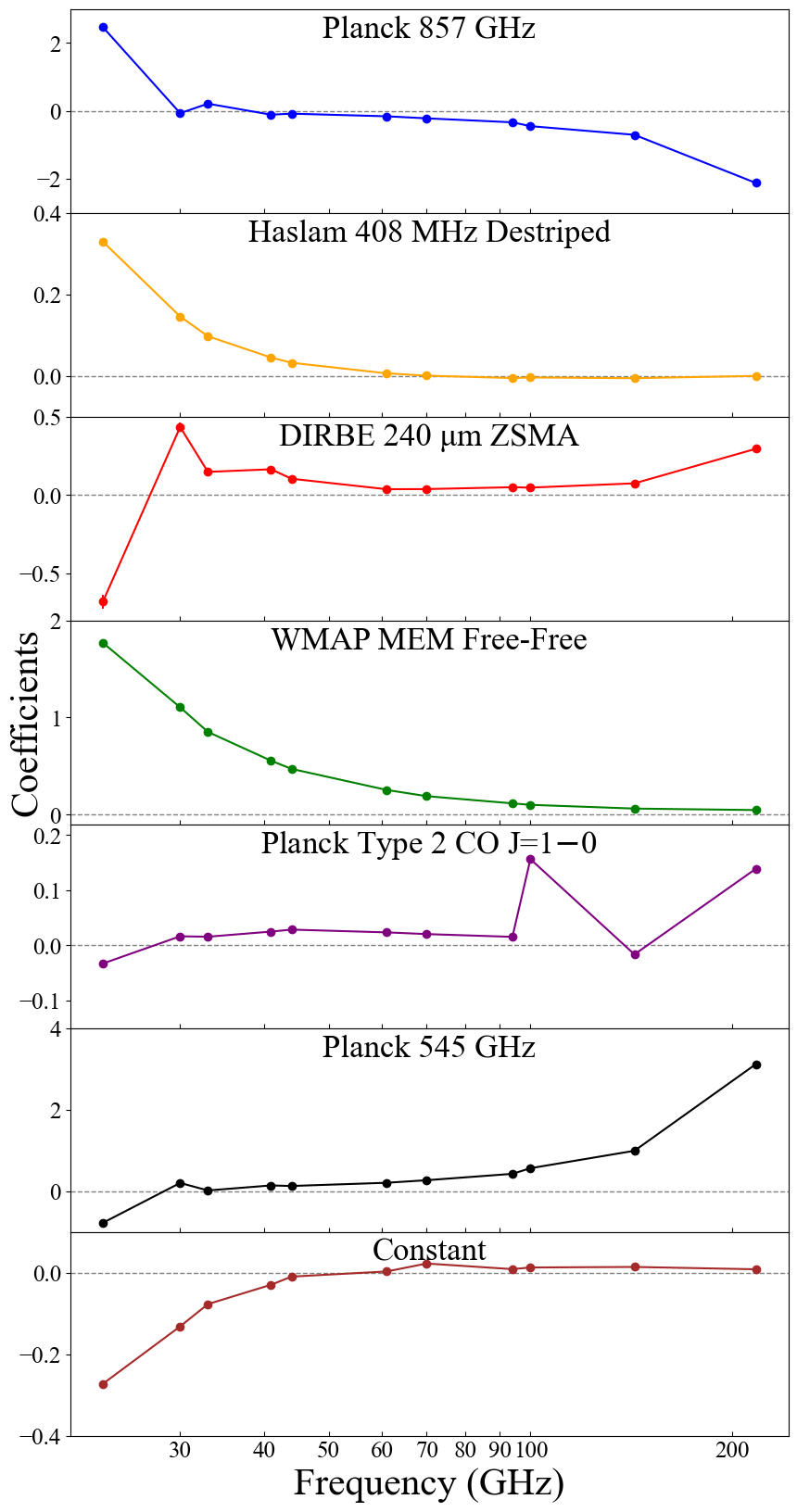}
    \caption{Template coefficient values for each frequency with arbitrary rescaling of the data for the final maps. No spectral index constraints are imposed, but spectral index results can be inferred from these fit results. 
    We find a free-free spectral index of $-2.13$ and a thermal dust index of 1.5.  Further details are presented in
    Section~\ref{sub:finalclean}. }
    \label{fig:prelim}
\end{figure}

We adopt the 1\% level of masked pixels, but note that there is nothing sensitively dependent on this exact choice. With this small sky area of masked pixels mode mixing is essentially no longer a problem in the analyses of the map. We experimented with inpainting, but we see no appreciable advantage to doing so.

\vspace{-0.25cm}
\subsection{Final Cleaning Process}
\label{sub:finalclean}
Since we attribute the pixels from the 1\% mask to errors in cleaning or systematics, we exclude them from the foreground fit. We fit the Galactic foreground emission templates within the $-10\degree < b <+10\degree$ strip with the $1\%$ mask applied. We execute a subsequent removal of the monopole and dipole across the full-sky with the $1\%$ mask applied.  The full-sky projections of the final best cleaned CMB maps are shown in Figure \ref{fig:fullskyr2}.

\begin{table*}[tbh]
    \centering
    \begin{tabular}{cccccccc}\\
        \hline
        Frequency & Haslam 408 & DIRBE 240 & WMAP MEM FF & Planck CO & Planck 545 & Planck 857 & Constant \\\hline
        23 GHz & 0.3291 & $-$0.6812 & 1.7969 & $-$0.0330 & $-$0.7793 & 2.4626 & $-$0.2729\\
        30 GHz & 0.1454 & 0.4330 & 1.1038 & 0.0165 & 0.1991 & $-$0.0702 & $-$0.1320 \\
        33 GHz & 0.0971 & 0.1469 & 0.8507 & 0.0157 & 0.0148 & 0.2092 & $-$0.0774\\
        41 GHz & 0.0447 & 0.1636 & 0.5541 & 0.0250 & 0.1379 & $-$0.1140 & $-$0.0298\\
        44 GHz & 0.0319 & 0.1032 & 0.4696 & 0.0286 & 0.1247 & $-$0.0846 & $-$0.0099\\
        61 GHz & 0.0059 & 0.0364 & 0.2527 & 0.0236 & 0.2033 & $-$0.1621 & 0.0027\\
        70 GHz & 2.4881e$-$4 & 0.0373 & 0.1886 & 0.0205 & 0.2660 & $-$0.2221 & 0.0222\\
        94 GHz & $-$0.0056 & 0.0491 & 0.1145 & 0.0155 & 0.4226 & $-$0.3390 & 0.0085\\
        100 GHz & $-$0.0042 & 0.0467 & 0.0993 & 0.1566 & 0.5605 & $-$0.4520 & 0.0126\\
        143 GHz & $-$0.0060 & 0.0738 & 0.0608 & $-$0.0165 & 0.9884 & $-$0.7072 & 0.0138\\
        217 GHz & $-$3.4304e$-$4 & 0.2941 & 0.0454 & 0.1390 & 3.1126 & $-$2.1233 & 0.0081\\
    \end{tabular}
    \caption{Template coefficient values from cleaning process fit as described in Equation \ref{equ:temp}. Column 1 designates the frequency $\nu$, columns 2$-$7 are the $a_{i,\nu}$ values associated with each rms-normalized template map $T_i$ given in the header, and the last column is $C_{\nu}$.}
    \label{tab:coeff_flipped}
\end{table*}

The template scaling coefficients derived in the final cleaning process are given in Table \ref{tab:coeff_flipped} and shown
graphically in Figure~\ref{fig:prelim}.  Coefficients for individual templates are plotted in a relative sense (arbitrary units) as a function of frequency from 23 to 217 GHz.  While spectral index constraints were not imposed, they can be inferred from these fit results.  The free-free spectral index derived from the fit coefficients for 23 - 217 GHz is $-2.13$, in close agreement with physical expectations, as discussed in the Introduction.  This is the template with least morphological
degeneracy with the other five so this result is a reassuring check on the method.  
The Haslam 408 MHz fits also appear to give a power-law. It is presumed to trace synchrotron, at least to some extent, but the spectral index derived from the fits at 23 - 44 GHz is $-3.65$, steeper than expected within
the Galactic plane \citep{quijote_plane:2023}. There are no templates that directly provide for synchrotron spectral index variations so decoherence of the morphology could play a role as could some absorption of synchrotron emission into other templates. The template morphology has similarities with other templates so an accurate synchrotron spectral index is not expected.  
We note that for $94 - 143$ GHz, the Haslam template contributes very little to the final foreground fitting solution.
The CO template is clearly used to represent CO at 100 and 217 GHz, but the fit indicates that it is
also used as a useful tracer for other mechanisms at lower frequencies. 
The constant value corrects for the differences in the monopole between the target maps. The decrease of the constant coefficient towards lower frequencies may be due to the strong signals and the decline in fit quality with decreasing frequency.

\begin{figure*}[hbt!]
    \includegraphics[width=7in]{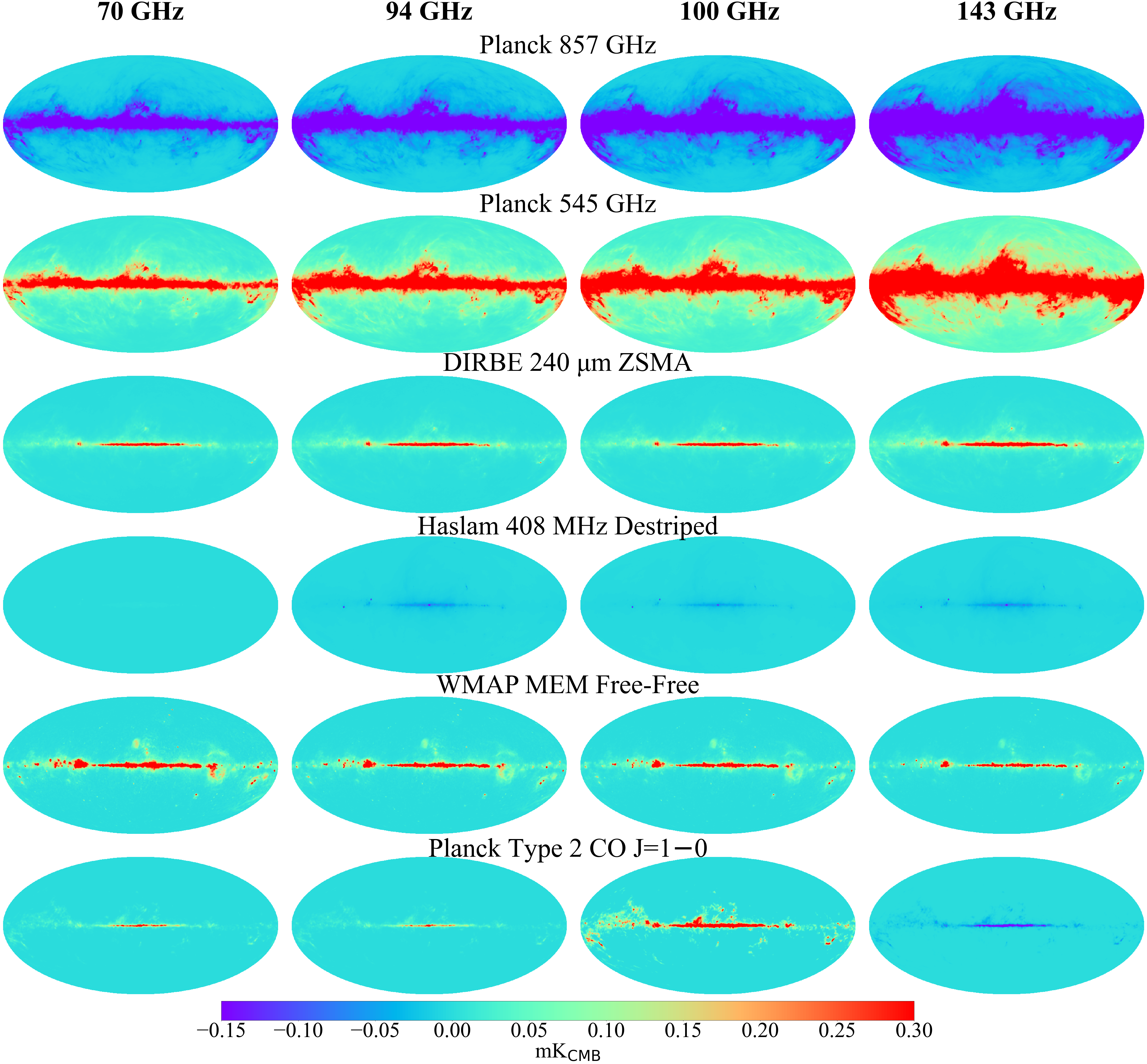}
    \caption{Full-sky maps of the templates with their as-fit amplitudes for the final cleaned CMB maps at (from left to right columns) 70 GHz, 94 GHz, 100 GHz, and 143 GHz. The foreground templates are scaled to the amplitudes as used in the foreground-cleaning fit. }
    \label{fig:24templates}
\end{figure*}

Contributions of each template in temperature units are illustrated in Figure \ref{fig:24templates}, which shows the templates multiplied by their as-fit amplitudes for each of the four frequencies of the cleanest CMB maps.
In this frequency range, both thermal and AME dust, along with free-free emission are the primary foregrounds.
Five of the six core foreground templates either directly represent dust emission (545 GHz, 857 GHz, DIRBE 240 $\mu$m)
or have some correlation with dust spatial morphology (Haslam, CO) in the Galactic plane.
The multiple templates with dust morphology serve to compensate for dust temperature variations along the plane
\citep{schlegel/finkbeiner/davis:1998, bennett/etal:2013, planck/10:2016}
and the lack of a single template that sufficiently traces AME.
Under the assumption that these five templates are being used as proxies for thermal dust emission at 94 - 143 GHz
(with the exception of the 100 GHz CO fit), we can derive an approximate thermal dust spectral index by
summing the contributions from these templates to produce a dust map at each of these frequencies, and then
fitting the rms spectrum for a power-law index.  This results in a spectral index $\beta_{\rm{dust}} =1.5$, which is
similar to the value of 1.4 we find for the same region for the Planck PR2 \texttt{Commander} model.

\subsection{Quality Analysis}
\label{sub:analysis}
\begin{figure*}
    \includegraphics[width=7in]{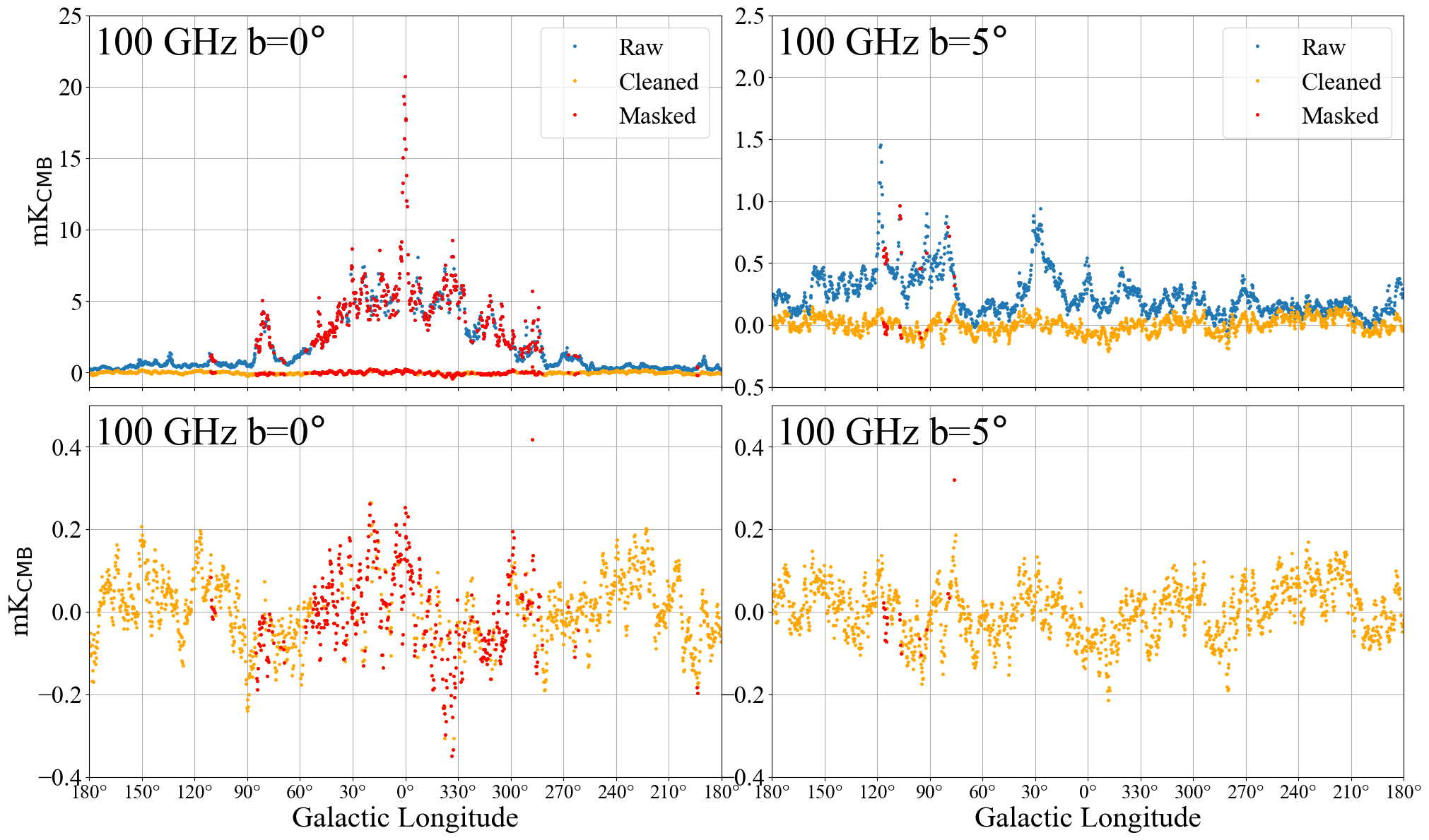}
    \caption{Slices in Galactic latitude of the 100 GHz raw and cleaned CMB maps. Top left: The $b=0\degree$ slice through the Galactic plane is shown both before and after cleaning. Although the entire slice reflects a substantial reduction of foreground levels, a high fraction of these solutions had an elevated standard deviation between the 70 GHz, 94 GHz, 100 GHz, and 143 GHz cleaned CMB maps and are thus included in the 1\% mask, as shown in red. The strong emission at the Galactic center is from CO. Bottom left: An expanded view of the cleaned solutions from the plot above. The fluctuation pattern is dominated by the CMB, not foreground residuals, especially for the unmasked points. 
    Top right: Same as the top left plot but for a $b=5\degree$ slice and the temperature scale has been expanded by a factor of 10. A few pixels fail the standard deviation test and are masked, but the vast majority of foreground-cleaned points reliably trace CMB fluctuations. Bottom right: An expanded view of the cleaned solutions from the plot above.} 
    \label{fig:slices}
\end{figure*}

Figure \ref{fig:slices} provides plots that illustrate the suppression of the foregrounds for the final maps. These plots are slices at fixed Galactic latitude at $b=0\degree$ and at $b=5\degree$. There are 1966 pixels in the 1\% mask and 438 (22.3\%) of these are in the $b=0\degree$ slice, reflecting an elevated deviation level between the CMB cleaned maps at 70, 94, 100, and 143 GHz. There are 119 masked pixels in a $2\degree$ slice and 20 in the $5\degree$ slice. As can be seen, the foregrounds are substantially reduced even in the masked pixels. Unmasked pixels reflect real CMB fluctuations to a fraction of the cosmic variance level. 

Figure \ref{fig:differencesr1} shows the difference maps from the initial cleaning over the full-sky. Following the final cleaning, we can quantify the foregrounds in the cleaned map differences, excluding 1\% of masked pixels. Specifically, the rms of the map differences for the 1\% mask are: $\sigma_{70,94}=6$ $\mu$K, $\sigma_{70,100}=4$ $\mu$K, $\sigma_{70,143}=5$ $\mu$K, $\sigma_{94,100}=6$ $\mu$K, $\sigma_{94,143}=6$ $\mu$K, $\sigma_{100,143}=2$ $\mu$K, which are all small compared to the 70 $\mu$K CMB rms, $< 0.8\%$ of the CMB variance, with no obvious spectral dependence. These rms differences include contributions from the noise quantified in Section \ref{sec:dataanalysis}.

The next best-cleaned map, which we have not used, is WMAP 61 GHz. It has both high noise and a factor of $\sim 3$ larger foreground residuals than our 4 best-cleaned maps. We judged that the benefit of 5 instead of 4 cleaned maps did not warrant its inclusion in our set of best-cleaned maps.

\begin{figure}[hbt!]
    \includegraphics[width=3.2in]{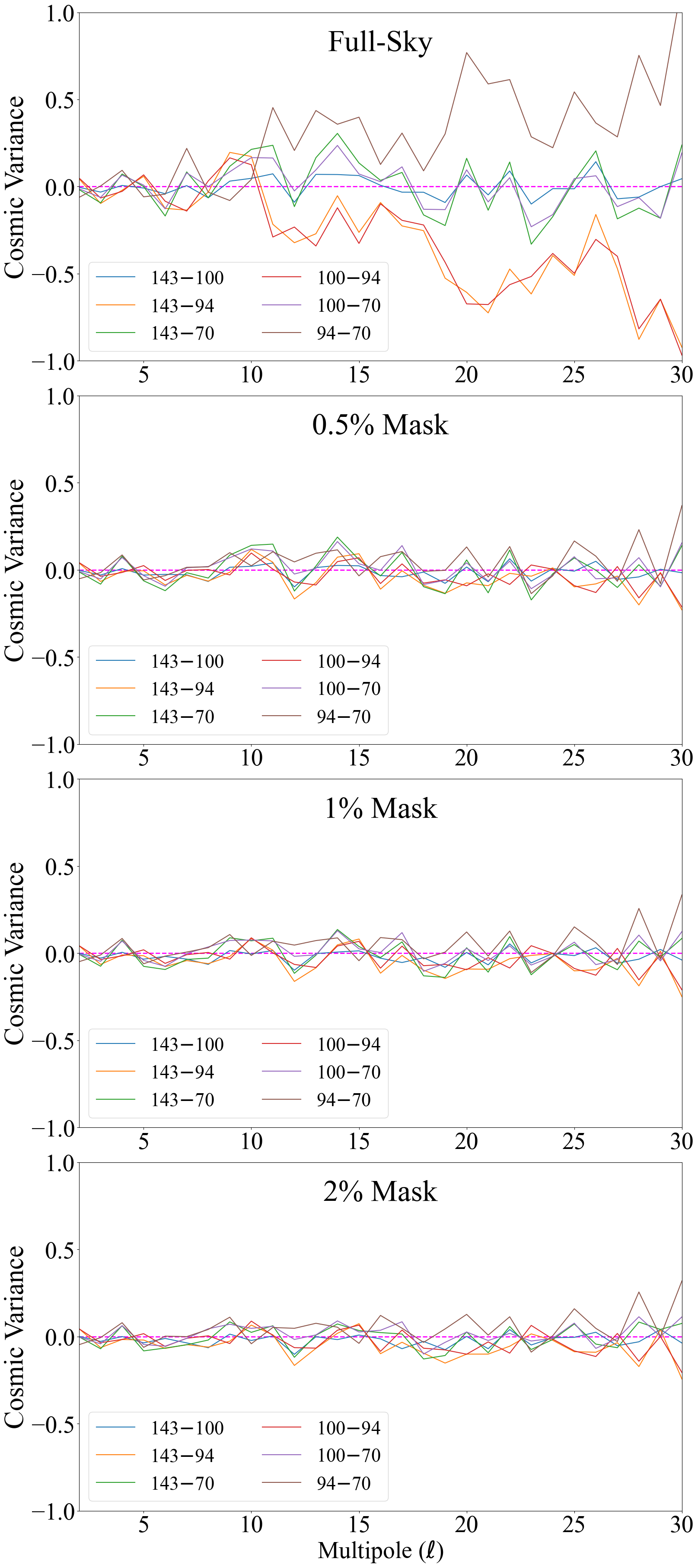}
    \caption{Six power spectra differences between pairs of the four CMB maps, in cosmic variance units, for cleaned maps with no mask, $0.5\%$, $1\%$, and $2\%$ masks applied. The power spectra differences are generally $<0.2$ of the cosmic variance and do not depend strongly on the choice of mask.}
    \label{fig:ps_diff_all}
\end{figure}

\begin{figure}[hbt!]
    \includegraphics[width=3.2in]{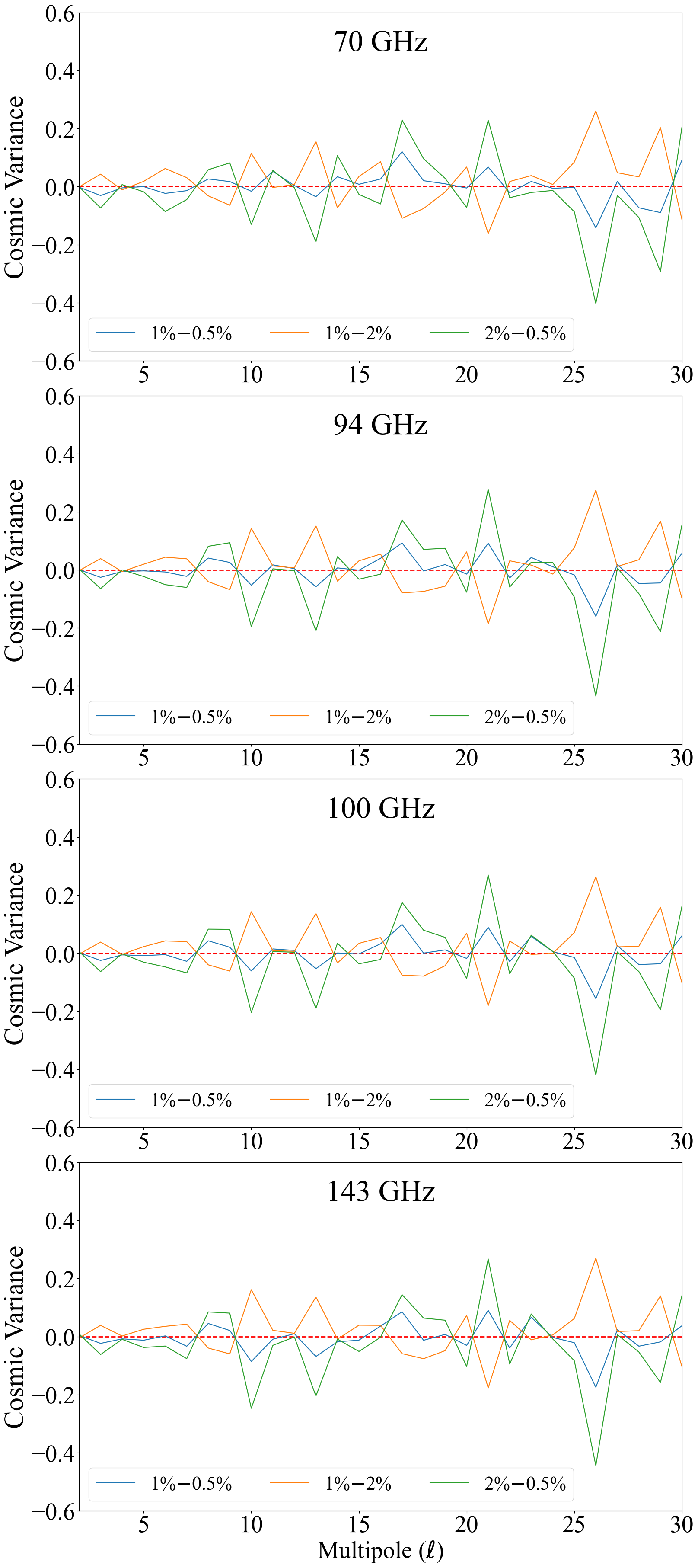}
    \caption{Power spectra differences between mask levels for each of the four cleaned CMB maps, in cosmic variance units. Together with Figure \ref{fig:ps_diff_all} we see that the power spectrum differences between the four frequencies and between mask levels are small compared to the CMB cosmic variance.}
    \label{fig:ps_diff_masks}
\end{figure}

Having arrived at four cleaned CMB maps and having deduced masks based on the greatest temperature standard deviation between the four maps, we now examine the power spectrum differences between maps.We offer two views of these differences. Figure \ref{fig:ps_diff_all} shows the power spectra differences between the 6 frequency pairs of the 4 CMB cleaned maps. Power spectra are shown for the full-sky, 0.5\%, 1\%, and 2\% masks. Figure \ref{fig:ps_diff_masks} shows the power spectrum differences between masking levels for each of the four CMB maps.

We draw two major conclusions from these power spectrum differences: (1) the power spectrum differences between the four CMB maps  are small, $<0.2$ of the CMB rms. This suggests that all four maps were well-cleaned and that they are dominated by CMB fluctuations, even in the Galactic plane; (2) while we think it is wise to mask some pixels that have a high rms between maps, the exact choice of masking level is not critical. 

As another check of the quality of the four cleaned maps, we compare the rms of four Planck PR3 cleaned maps, with the Planck common mask applied, with the four cleaned maps produced here over the same mask. All maps were smoothed to a common $1^{\circ}$ resolution. We find the rms of the four cleaned maps in this paper agree to within 1\% percent with each of the SMICA, NILC, SEVEM, and Commander rms over the common mask. Although our templates were fit over the brightest portions of the sky ($|b|\leq10^{\circ}$), there is no evidence for over- or under-subtraction of foregrounds at the higher latitudes.

If only a single map is desired, our cleaned 100 GHz map with 1\% masking would be a reasonable choice since it has low noise and low residual foregrounds. However, we view having four independent maps of the CMB a significant advantage for statistical analyses. In general, we recommend use of all four of our maps, as we have done in the two companion papers \citep{Nofi:2025b, herold2025}.

\section{Conclusions}
\label{sec:conclusions}

This paper was motivated by an interest in obtaining observational
maps of the CMB temperature anisotropy, cleaned of Galactic foreground emission
over as much of the full sky as possible, including the Galactic plane, without use of inpainting. Our emphasis is on the large angular scales $2 \leq \ell < 30$, using maps at $1^\circ$ resolution.

To clean foregrounds we fit a linear combination of multiple foreground templates 
to a target frequency map.  This fit is used to remove Galactic foregrounds independently of the emission mechanisms 
present in the target map. While we attempt to represent the
morphologies of major physical components in the selection of templates, we are not concerned with component
separation or modeling the sky component parameters. We
rely instead on the morphologies of multiple template maps to combine in whatever way minimizes the overall
foreground contamination.

We started with a set of 11 target frequency maps ($23 - 217$ GHz) from the WMAP 9-yr and Planck PR3 data releases,
and a set of 17 candidate templates (Table~\ref{tab:templates}).   We summarize our findings as follows:

(1)  From the initial set of templates, a core set of only six templates was used in our final fits.  The addition of
other templates did not significantly improve the quality of the foreground removal. A combination of these six templates at each target frequency band
appears to have adequately provided the ability to represent the majority of the
complex Galactic emission.

Furthermore, the six chosen templates have an extremely low probability of subtracting the CMB signal. 
In Section \ref{datatempcriteria} the third criterion stated the template maps must be strongly foreground-dominated. The DIRBE 240 $\mu$m and Haslam 408 MHz templates have essentially zero CMB contribution. In Section \ref{sub:finaltemp} we have discussed the low probability that there is a significant CMB signal contaminating the free-free and CO templates. We have verified that the CMB contributions from the scaled 545 and 857 GHz templates are limited to a fraction of a percent, so can be safely ignored.

(2)  Differences between the cleaned maps revealed notable foreground residuals at frequencies $23 - 61$ GHz and 217 GHz. The best cleaned maps were at 70, 94, 100 and 143 GHz. 
Additionally, the region around SgrA
showed clear variability between the WMAP and Planck observational eras.  We excluded this region 
(0.02\% of all pixels) from  further analysis.

(3) The cleaned 100 and 143 GHz maps
showed a low-level, but distinctive, zodiacal light residual in the form of two parallel bands 
roughly $10^\circ$ above and below the ecliptic.  This feature is also seen in pair differences between some of the
Planck CMB map 
products (see e.g. Figure 7 in \citealt{planck/04:2018}).  We corrected the small over-subtraction in 100 and 143 GHz using a
template (Section~\ref{sub:dataprep}), and used these corrected maps in subsequent fitting and analysis.  

(4) Additional quality tests of the four best-cleaned maps (histograms, map differences between frequencies,
power spectrum differences) indicated a need for additional masking, particularly close to the Galactic plane.
We formed masks excluding 0.5\%, 1.0\% and 2\% of pixels (including the SgrA region); details of the masking
algorithm are described in Section~\ref{sub:masks}. The masking threshold itself is not critical, and we find no sufficient statistical benefit to masking more than 1\% of the pixels.

(5) The final 70, 94, 100 and 143 CMB maps were cleaned over 99\% of the sky.  Power spectrum differences,
histograms, and map differences (Sections~\ref{sub:finalclean}, \ref{sub:analysis}) all indicate that the maps are strongly dominated by CMB.  
Although differences between the four CMB-cleaned maps still have 
foreground-related residuals close to the Galactic plane, these are at a level well below the CMB cosmic variance for pixels not excluded by the mask. The four maps are indistinguishable to within a fraction of their cosmic variance levels.

(6) While no spectral constraints were imposed in our fits, physically reasonable spectral index values 
for certain foreground components are recovered from the template coefficients.  This includes a free-free spectral 
index of $-2.13$, and a thermal dust spectral index of 1.5.  This provides some confidence in our cleaning solution.
However, our fits did not remove foregrounds adequately for bands other than the four frequencies we highlight. 
Several factors might contribute to this: (a) non-orthogonality (degeneracy) between template morphologies within the plane 
(b) the inability of a single 408 MHz template to characterize spatial spectral variations in the synchrotron emission (decoherence); (c) lack of any fully predictive spatial template for AME, and (d) the greater challenge in fitting foregrounds at frequencies where the foreground signal is high and less forgiving of removal imperfections.

(7) We present a set of four CMB maps for low-$\ell$ studies, maximizing sky coverage (99\%) for CMB information while minimizing adverse effects of masking, such as mode-mixing when analyzing the map data.   In this sense, our maps differ from the more general derivations of the WMAP ILC and
the Planck {\texttt{Commander}}, \texttt{SMICA}, \texttt{NILC} and \texttt{SEVEM} CMB estimates, which were recommended for use with a 
conservative (larger) mask about the Galactic plane.  A substantial benefit of this work is the production of  four independent CMB maps based on four frequencies that were cleaned. The inter-comparison of these maps provides an estimate of uncertainties as we have shown in Figures \ref{fig:stdvhist}, \ref{fig:fullskyr2}, \ref{fig:ps_diff_all}, and \ref{fig:ps_diff_masks}.

The use of four CMB maps with independent noise enables robust estimation of statistical quantities and their uncertainties. As we show in the companion paper \cite{Nofi:2025b}, the four independent CMB maps enables the computation of multi-frequency cross power spectra. For other statistical analyses, we recommend a similar approach be used that takes advantage of all four maps. A second companion paper \cite{herold2025}, focuses on the characterization of CMB anomalies. This work illustrates the value of four independent CMB map solutions as a measure of statistical uncertainty.

In addition to illustrating the advantages of using four independent CMB maps, these companion papers provide strong evidence that the four maps are well-cleaned and statistically consistent with one another. 

The four independent CMB maps (``External Linear Combination (ELC) maps") and code to reproduce the results in this work are available in a Zenodo repository \citep{Nofi:19709972}.

\vspace*{0.25in}

This research was supported by NASA grants 80NSSC23K0475, 80NSSC24K0625, and 80NSSC25K7518.
We acknowledge the use of the Legacy Archive for Microwave Background Data Analysis (LAMBDA), part of the High Energy Astrophysics Science Archive Center (HEASARC). HEASARC/LAMBDA is a service of the Astrophysics Science Division at the NASA Goddard Space Flight Center.
We also acknowledge use of the \textit{Planck} Legacy Archive. \textit{Planck} is an ESA science mission with instruments and contributions 
directly funded by ESA Member States, NASA, and Canada.

\software{numpy \citep{harris/etal:2020}, scipy \citep{virtanen/etal:2020}, matplotlib \citep{hunter:2007}, astropy \citep{astropy2022}, HEALPix/healpy \citep{gorski/etal:2005, healpy:2019}}

\clearpage
\newpage

\bibliography{gforegrounds,software,wmap}
\bibliographystyle{aasjournal}

\end{document}